%% file: main.tex
\documentclass{article}[12pt]
\usepackage{amssymb}
\usepackage{geometry}
\usepackage{graphicx}
\usepackage{graphics}
\usepackage{mathtools}
\usepackage{amsmath}
\usepackage[utf8]{inputenc}
\usepackage{lettrine}

\title{Probability density function of the unbalanced impulse in Langevin’s theory of Brownian motion}
\author{
  Ayanabha De \\
  Department of Physics, \\
  Indian Institute of Technology Madras, Chennai, India \\
  \texttt{ph21c005@smail.iitm.ac.in}
}

\begin{document}

\maketitle

\begin{abstract}
    
This paper attempts to find a probability distribution for the white noise (rapidly fluctuating unbalanced force) in the Langevin's Equation. Unbalanced force is the resultant impulse provided to the brownian particle by the colliding fluid molecules. Therefore, a probability distribution of the speed of the particles after each impact will have the same probability distribution of the white noise. Such a distribution is discovered in this work by constructing a simple model based on thermal molecules colliding with the particle from all directions. The molecules obey Maxwell-Boltzmann speed distribution law. At low temperatures, for bigger brownian particles, existence of some non-random distribution for the unbalanced impulse, in itself is an interesting result. The distribution takes a "near" half gaussian form at these limits. At high temperatures, for small brownian particles(e.g: pollen grains), the distribution is shown to approach uniform distribution, and hence consistent with bulk of well established theoretical assumptions and experimental results in the literature that claims the unbalanced force to be a random white noise.

\end{abstract}

\input{sections/introduction}

\input{sections/model}

\input{sections/PDFofspeed}

\input{sections/results}

\input{sections/conclusions}

\input{sections/Acknowledgement}

\bibliographystyle{plain}
\bibliography{bibfile}

\end{document}

%% file: sections/introduction.tex
\section{Introduction}

Brownian motion is an irregular motion of particles suspended in a fluid. Some of it's important properties as found experimentally are as follows,
\begin{itemize}
    \item It is very random and it includes translations and rotations of the particle.
    \item Smaller particles exhibit more active motion.
    \item More active motion is seen in fluids of lesser viscosity.
    \item More active motion is seen in fluids with higher equilibrium temperature.
    \item It doesn't depend on the  composition of the fluid.
    \item The motion is eternal.
\end{itemize}
Einstein \cite{einstein} had given the first explanation of the phenomena and Perrin confirmed it \cite{perrin}. Langevin gave a dynamical theory of Brownian motion\cite{Coffey} by writing an equation of motion for the position of the Brownian particle ($x$) according to Newton's laws under the assumptions that the particle experiences two types of forces,
\begin{itemize}
    \item Viscous drag force, $-\zeta \frac{dx}{dt}$ where $\zeta$ is the coefficient of viscosity and $t$ represents time.
    \item An unbalanced impact force $F(t)$ due to the impulse imparted by the colliding fluid molecules. 
\end{itemize}
The Langevin's equation is given as,
\begin{equation}
    m\frac{d^2 x}{dt^2} = -\zeta \frac{dx}{dt} + F(t)
\end{equation}
The viscous force is analogous to a friction force which tends to slow down any motion in the fluid. But the eternal Brownian motion is the resultant effect of the momentum imparted by each of the colliding molecules. This unbalanced force (known as white noise in modern literature) is the subject of interest in this paper. The white noise conveniently averages to zero and gives an useful solution\cite{Coffey} to the Langevin's equation. This method brought light onto Stochastic processes in physics which has now become an entire research area on its own.

\par

This article, however, asks an interesting question. The molecules colliding with the Brownian particle themselves obey some energy distribution law(e.g: Maxwell-Boltzmann distribution for classical thermal fluid). Shouldn't this fact be able to lead us to the exact probability distribution law for the energy(or momentum) it imparts on the Brownian particle ? This paper is my curious and successful attempt to find such a probability density function for the unbalanced impulse which is also consistent with observations. The results obtained show little promise to have any significant effect on the extensively studied field of Brownian motion. However, it might interest some low temperature experimental physicist to validate my results at their leisure. I have published this paper mostly because it is an interesting academic exploration.

%% file: sections/model.tex
\section{Model}

The main concept that drives this work is that the molecules that are colliding and imparting the momentum to the brownian particle, themselves follow a speed distribution law, and hence it is expected that the resultant speed of the particle too will obey some speed distribution law. To quantify this conceptual realization, the following set up is constructed.

\begin{figure}[httb!]
    \centering
    \includegraphics[width=2in]{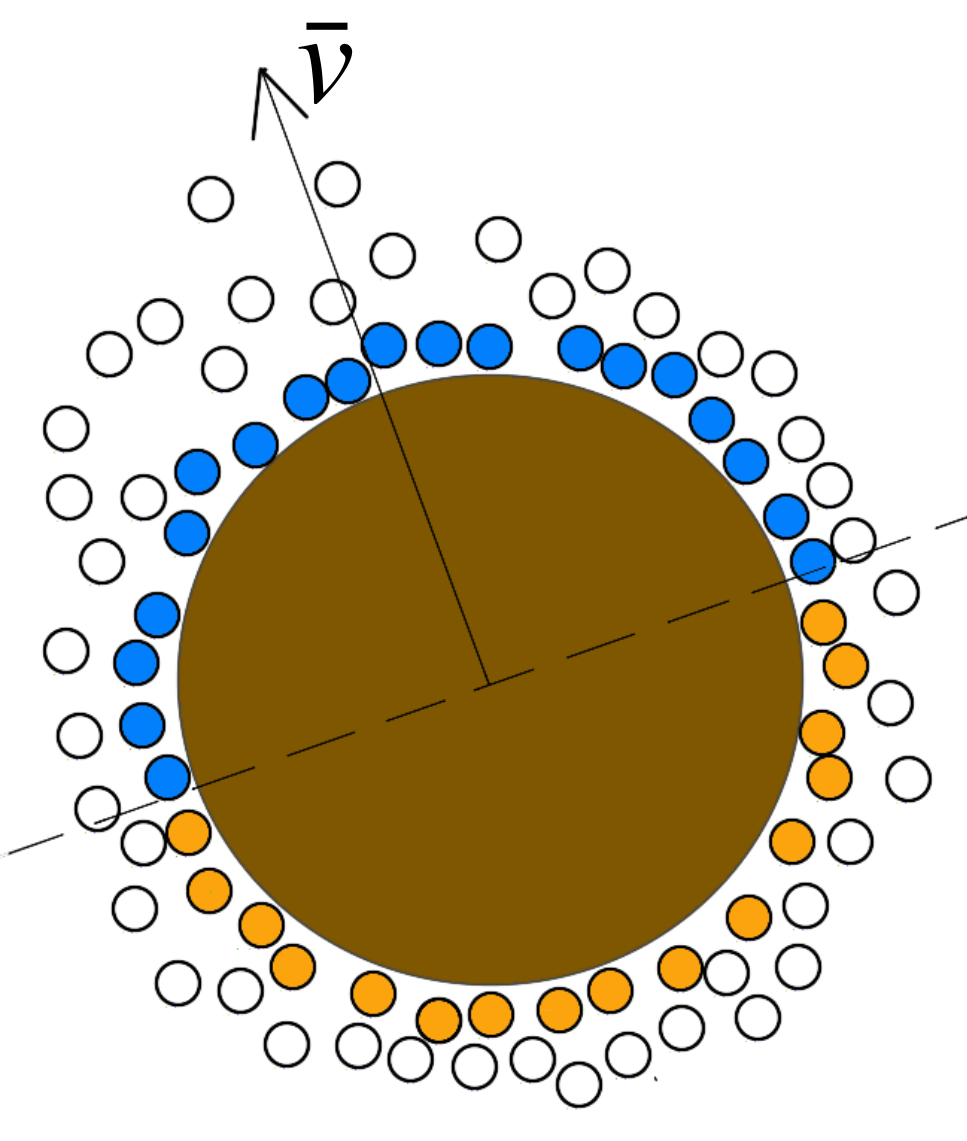}
    \caption{The vector $\bar v$ represents the velocity of the brownian particle right after the collision with the fluid molecules nearby. The dotted line is a diameter perpendicular to the velocity vector drawn from the origin. The dotted line divides the collding molecules into orange and blue half. This construction is important for the proceeding formulation.}
    \label{particle}
\end{figure}

The brownian particle is studied at a particular instant right after a collision with the nearest fluid molecules. To obtain the probability distribution of the resultant speed of the particle, we shall take the ensemble average of the same situation. 
\par
Figure \ref{particle} is a 2 dimensional realisation of a particle suspended in a midst randomly moving molecules which compose the fluid. There are very many fluid molecules in total, but here for a neat depiction of the phenomenon, I have shown a smaller number of molecules. At the instant which this sketch depicts, only the molecules coloured blue and orange, are colliding with the particle. At this instant, the Brownian particle is at rest with respect to the origin which sits at the intersection of $\bar v$ and the diameter drawn perpendicular to it. Infinitesimal time after this instant, the particle is going to start moving along $\bar v$. This represents the response of the particle after to the impulse from the colliding molecules.   

The component of momentum perpendicular to $\bar v$ imparted on the particle by the blue molecules must cancel out the ones imparted by the orange molecules. But along the $\bar v$, the parallel components are unbalanced. Thus the unbalanced imparted impulse is acting along the $\bar v$ through the centre of the particle. It can be visualised that there is a push of $N_1mC_1$ and $N_2 mC_2$ by the orange and the blue molecules respectively. Here $m$ is the molecular mass of the fluid, $N_1$ and $N_2$ are the number of orange and blue molecules respectively. Now, it is assumed that the colliding molecules obeys Maxwell-Boltzmann speed distribution law. This will be discussed in detail soon. The subset of molecules labelled in orange and blue must then also obey the same distribution law if $N$ is large enough. $C_1$ and $C_2$ are the averages of the speed(along $\bar v$) distributions of the orange and blue subsets. Then an $n^{th}$ subset of the fluid will push the brownian particle with a momentum $m(N_n C_n)$ along $\bar v$.

\subsection{Assumptions}
Now before proceeding further to formulate the probability distribution, I shall lay down the primary assumptions made in the formulation.
\begin{itemize}
    \item The fluid is at thermal equilibrium and obeys Maxwell-Boltzmann speed distribution law. My model is applicable even for fluids obeying other types of speed distribution. But I have made this assumption for simplifying calculation and also because most fluids in the laboratory would obey this law.
    
    \item The size of the particle is large enough compared to molecular dimensions, such that number of molecules colliding with the particle at a specific time is large enough to precisely obey a Maxwell-Boltzmann speed distribution curve.

    \item Though the calculations will generally account for any shape of the particle, I have assumed the brownian particle to have a spherical shape for simplifying calculations.
    
    \item Also for simplifying calculations, it will be assumed from hereon that the number of colliding molecules is equal(=$N$) in each of the subset of molecules. This assumption is reasonable at high number of subset molecules(i.e: big size of brownian particle and thermodynamic limit).
\end{itemize}
\subsection{Speed distribution law of the fluid}
The fluid at thermal equilibrium obeys Maxwell-Boltzmann distribution law. Now from this huge number of molecules, if we create smaller subsets of molecules, they too would obey the same distribution law. But, there will be some statistical fluctuations which will reflect in their averages. So, suppose the fluid is at equilibrium temperature $T$. If the fluid is divided into $i$ number of subsets, each of the subset will obey have some temperature(analogous average to the average speeds of the molecules) that is different from $T$. That is, mean speed of each of the subsets will fluctuate from the mean of the entire fluid. But the net fluctuation will always add up to exactly zero so that the thermal equilibrium of the fluid is maintained. That is, If the fluctuation for $k^{th}$ subset is $dT_k$, then
$$
    \sum_{k=1}^i dT_k = 0
$$
Higher the number of subsets, greater fluctuations can occur. The average speed of each subset, $C_k$ corresponding to the fluctuated temperature, $T + dT_k$ will be central to our proceeding formulation and as we will see, it will be the reason for the "imbalance" in the unbalanced force.

%% file: sections/PDFofspeed.tex

\section{Probability Density Function of Brownian particle speed}
Let the resultant speed of the brownian particle immediately after the collision depicted in Figure {\ref{particle}} be $v = |\bar v|$. Now suppose the collisions are elastic. Then given that the orange subset has average speed $C_1$ and that the blue subset has average speed $C_2$, the conservation of momentum provides us with a relation between $v$ and $C_1$,$C_2$ as,
$$
    C_1 - C_2 = \frac{Mv}{mN}
$$
But for any general type of collisions, the relation would look like,
\begin{equation}
     C_1 - C_2 = f(v)
     \label{eq1}
\end{equation}

Probability that the particle will have the resultant speed $v$ given that the blue subset has average speed $C_2$ is, 

$$ p(v) = p({C}_2 + f(v))p({C}_2)$$\\

But the value of $C_2$ can have any value. For all practical purposes we can take it's range to be from $0$ to $\infty$. We then need to integrate the above expression to obtain the total probability of the brownian particle to have a resultant speed $v$.

\begin{equation}
\displaystyle  P(v) = \frac{\int\limits_{C_i =0}^{\infty} p(C_i + f(v))p(C_i)dn_{C_i}}{\int\limits^\infty_{C_i =0} dn_{C_i}}
\label{eq2}
\end{equation}
where $p$ is the probability function that we would derive in subsequent sections and $dn_{C_i}$ is the number of subsets having average speed between $C_i$ and $C_i + dC_i$. Here the probability of each sample to have an average speed of $C_i$, is given by a gaussian(or normal) distribution according to Central Limit theorem(CLT). Normal distribution and CLT will be discussed briefly in the following section 3.1 to keep this article self-sufficient. A reader well versed with these topics can skip to the next section 3.2.

\subsection{Normal Distribution and Central Limit Theorem}

Suppose there a set of $n$ random variables,
$$x_1,x_2,......x_n$$
From this arbitrarily select a subset of $r$ random variables,
$$x_i,x_{i+1},......x_{i+r-1}$$
Then the average of the subset, $\displaystyle \frac{\sum_{i}^{i+r-1}x_k}{r}$ approximately follows a normal distribution with mean $r\mu$ and standard deviation $\sigma \sqrt{r}$, where $\mu$ and $\sigma$ are the mean and standard deviation of the distribution obeyed by the entire set of $n$ random variables. This is called the central limit theorem. It is considered one of the most important result of statistics \cite{Sheldon}. 

Normal distribution \cite{Sheldon} is a distribution function for random variable $x$, whose probability density function is given as,
$$\displaystyle N(x) = \frac{1}{\sqrt{2\pi}\sigma}e^{-{(x-\mu/\sqrt{2}\sigma)^2}}$$
If we change variable to $z = \frac{x - \mu}{\sigma}$, which is known as the standard normal variable, we obtain the following PDF in terms of the standard normal variable,
\begin{equation}
    \phi(z) = \frac{1}{\sqrt{2\pi}}e^{-{z^2}/2}
    \label{eq3}
\end{equation}
Probability of a certain random variable requires one to integrate this density function.
\subsection{Speed distribution}
For a large number of subsets, the average speed of subsets can be well approximated to obey a normal distribution according CLT. This normal distribution would naturally extend from $-\infty$ to $\infty$. But speeds can only have positive values. Therefore it had to be normalized for limits of $C_i$ from $0$ to $\infty$. This gives limits of $z_i(C_i)$ to be from $-\frac{\mu}{\sigma}$ and $z_i(C_i + f(v)) = z_i + F(v)$ where $F(v) = \frac{\sqrt{N}f(v)}{\sigma}$. Equation \ref{eq2} then converts to,
$$
    \displaystyle  P(v) = \frac{\int\limits_{z_i =-\mu/\sigma}^{\infty} p(z_i + F(v))p(z_i)dn_{z_i}}{\int\limits^\infty_{z_i =-\mu/\sigma} dn_{z_i}}
$$
$dn_{z_i} = n p(z_i)$ as it is the number of subsets having the average speed $C_i$. $\displaystyle \int\limits^\infty_{z_i =-\mu/\sigma} dn_{z_i}$ must be equal to $n$. Finding the exact form of probability distribution would require integrating the normal density function. That would be a difficult job. Rather, the probability density function would satisfy the aim of the paper and allow an analytical solution. Thus PDF of the speed of the brownian particle will be,
\begin{equation}
    \Phi(v) = \int\limits_{z_i =-\mu/\sigma}^{\infty} \phi(z_i + F(v))(\phi(z_i))^2 d{z_i}
    \label{eqn4}
\end{equation}

The renormalised expressions for $\phi(z)$ are,
$$\phi(z_i) = \left(\frac{1}{2}+\frac{1}{2}erf\left(\frac{\mu}{\sqrt{2}\sigma}\right)\right)^{-1}\frac{1}{\sqrt{2\pi}}e^{-z_i^2/2}$$
and
$$\phi(z_i + F) = \left(\frac{1}{2}+\frac{1}{2}erf\left(\frac{1}{\sqrt{2}}\left[\frac{\mu}{\sigma}+F\right]\right)\right)^{-1}\frac{1}{\sqrt{2\pi}}e^{-(z_i+F)^2/2}$$
where $erf(x)$ is the error function.

\begin{figure}[httb!]
    \centering
    \includegraphics[width=3.8in]{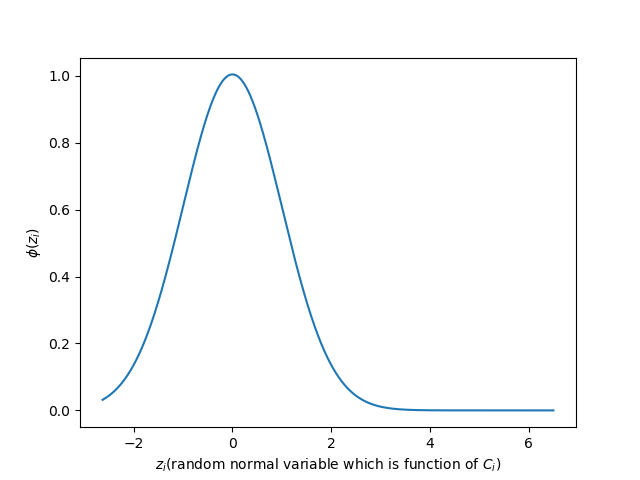}
    \caption{For a specific value of $\mu/\sigma = 2.63$, the normalised probability density for average speed of the subsets has been plotted. Note that it is an asymmetric Gaussian because the lower bound of the standard normal variable is $-2.63$ rather than $-\infty$.}
    \label{density}
\end{figure}

Substituting these values in Equation \ref{eqn4} and integrating we would obtain $\Phi(v)$. To make the resultant expression neat, 
$$A = \left(\frac{1}{2}+\frac{1}{2}erf\left(\frac{\mu}{\sqrt{2}\sigma}\right)\right)^{-1}$$
and 
$$A' = \left(\frac{1}{2}+\frac{1}{2}erf\left(\frac{1}{\sqrt{2}}\left[\frac{\mu}{\sigma}+F\right]\right)\right)^{-1}$$
Then on carrying out the integration in Equation \ref{eqn4}, we shall obtain,
\begin{equation}
    \Phi(v) = \frac{\alpha \sqrt{\pi}}{2}\left[1 + erf(\beta)\right] e^{-F(v)^2/5}
    \label{eqn5}
\end{equation}
 where $\displaystyle \alpha = \frac{2}{\sqrt{5}}\frac{A^2A'}{(2\pi)^{3/2}}$ and $\displaystyle \beta = \frac{\sqrt{5}}{2}\frac{\mu}{\sigma} + \frac{2F(v)}{\sqrt{5}}$
 \par
 In the whole calculation nowhere was value of $v$ constrained. So it ranges from $-\infty$ to $\infty$. But again, it is the speed of the brownian particle and therefore cannot have negative values. So that means $\Phi(v)$, when bounded from $v = 0$ to $\infty$, would no longer remain normalised. But the non-normalised form is enough for qualitative discussion that suffices the aim of the paper. Further, normalisation involves a pretty non-trivial integration which I leave to the reader's interest to carry out. Thus Equation \ref{eqn5} is the probability density function \footnote{It is important to note that this is a \textbf{non-normalised} probability density function for speed(to find the distribution function one needs to normalise it by integrating this PDF from $0$ to $\infty$) of the Brownian particle.} of the resultant speed of the brownian particle bounded between $0$ to $\infty$.

%% file: sections/results.tex
\section{Results and Discussions}
Now that I have the analytical form of the speed distribution function of the brownian particle, it is to be shown how it relates to the distribution of the unbalanced impulse.

\subsection{Distribution density of the unbalanced impulse}
For a given value of $v$, momentum of the brownian particle is $Mv$, where $M$ is the mass of the particle. And this itself is the impulse imparted by the molecules on the particle. Equation \ref{eqn5} is written in terms of $F(v)$ and 
$$F(v) = \frac{\sqrt{N}f(v)}{\sigma}$$
For a specific case of elastic collision, 
$$f(v) = \frac{Mv}{m}$$
Hence, for any specific type of collision, the PDF for impulse ($Mv$) will be same as the PDF of speed of the brownian particle ($v$) but scaled by some factor depending on the type of collision. Therefore, to qualitatively understand the probability distribution of the impulse, use of non-normalised PDF for $v$ in Equation $\ref{eqn5}$ will suffice. In the subsequent subsections, it is discussed how the PDF depends on temperature of fluid, size of the particle, and composition of the fluid.
\par
Hereon we are gonna make use of the assumption that the fluid obeys Maxwell Boltzmann speed distribution law and use it's corresponding values of mean and standard deviation.
\begin{equation}
    \mu = \sqrt{\frac{8RT}{\pi m}}
\end{equation}
and
\begin{equation}
    \sigma = \sqrt{\left(\frac{3\pi - 8}{\pi}\right) \frac{RT}{m}}
\end{equation}
The terms where mean and standard deviation appears as ratios, do not contribute to the dependence with our parameters of interest. Only $F(v)$ which solely depends on the standard deviation, causes the correlation between the PDF and the fluid parameters.
\begin{equation}
    F(v) = \frac{f(v) \sqrt{N}}{\sqrt{\left(\frac{3\pi - 8}{\pi}\right) \frac{RT}{m}}}
\end{equation}
Dependence of Brownian motion on $N$ is related to the dependence on size of the particle. And it is directly evident now that the motion depends on temperature $T$ and composition of the fluid $m$.
\subsection{Variation with size of the brownian particle}
Greater the size of the brownian particle, more number of molecules are going to collide with it, and hence higher will be the number of molecules per subset which is $N$. Therefore for larger particle size, $N$ would be higher.
\begin{figure}[httb!]
    \centering
    \includegraphics[width=3in]{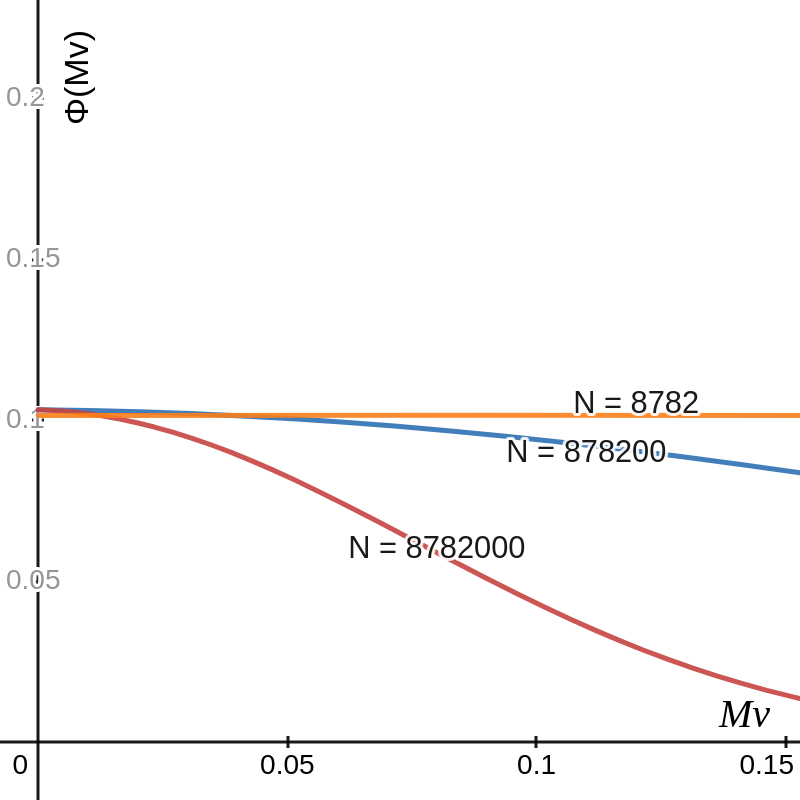}
    \caption{For constant temperature of 300K and molecular mass of 18g/mol, $\Phi(Mv)$ has been plotted against $Mv$ for different values of N, which corresponds to different size of the Brownian particle. Momentum axis is in SI units}
    \label{size}
\end{figure}

As evident from Figure{\ref{size}} that smaller the particle, lesser the number of molecules colliding with it, and hence higher momentum is imparted to the particle, that is "more active motion". Thus it is consistent with the size property listed in section 1.
\\
I have considered the Perrin's experiment \cite{perrin} where he took a suspended particle of size $R = 0.53 \mu m$ in water whose molar mass is 18. Radius of water molecules can be taken to be $r = 0.1925 nm$. I have assumed the experiment was performed in a room temperature of around $300K$. If one has to fit the water molecules most efficiently on the surface of the brownian article, then the number of molecules that can be fitted is,
$$\frac{4\pi R^2}{\pi r^2} \approx 3.1 \times 10^7$$
But water is a liquid, and not a solid crystal. Also among the liquids, it has relatively low density. This is why I have reasoned that the number of molecules actually colliding would be of the order which is only a small fraction of the most efficient lattice packing. I have approximated this fraction to be the ration of density of water to that of diamond. This is equal to $1/3530$. From this I obtain the value of $N$ for Perrin's experiment to be $8782$. Therefore the orange curve in the Figure $\ref{size}$ represents the probability distribution curve in Perrin's experimental setup. It is completely an uniform distribution. Langevin's approximation of completely random unbalanced force is realised here. 
\subsection{Variation with temperature of the fluid}
Variation of the probability distribution density with temperature is shown in the following plot.
\begin{figure}[httb!]
    \centering
    \includegraphics[width=3in]{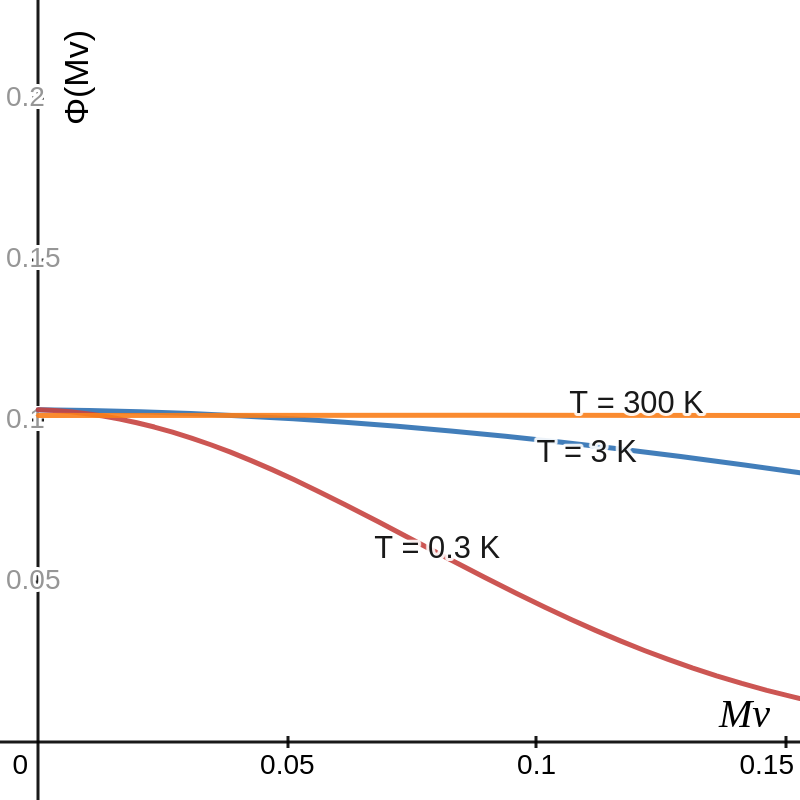}
    \caption{For constant size(N = 8482) and molecular mass of 18g/mol(again the parameters for Perrin's experiment), $\Phi(Mv)$ has been plotted against $Mv$ for different values of equilibrium temperature of the fluid.  Momentum axis is in SI units}
    \label{temp}
\end{figure}
From Figure $\ref{temp}$, it is clearly evident that for higher temperature, the motion becomes more active, as higher velocities become more probable. This again makes it consistent with the observed properties of Brownian motion. Again here, the orange curve represents Perrin's experimental setup.

\subsection{Dependence on composition of the fluid}
Composition of the fluid is characterised by the molar mass of the molecules that composes it. The following plot shows the variation of the PDF with changing molar mass $m$.
\begin{figure}[httb!]
    \centering
    \includegraphics[width=3in]{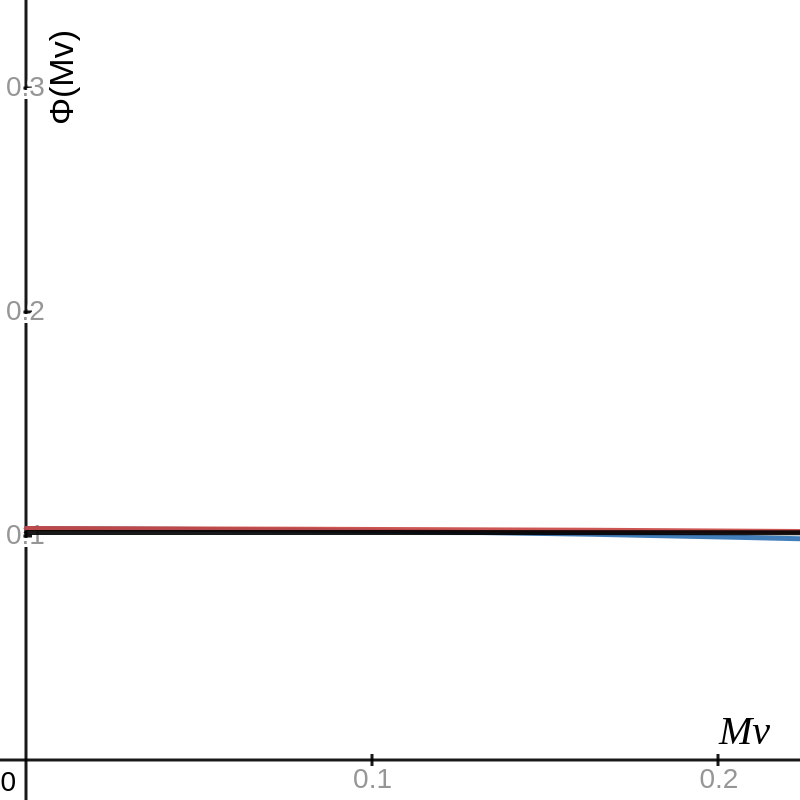}
    \caption{PDF has been plotted for constant temperature of $300K$ and size(N = 8482), but different values of molar mass. Plots are for $m = 18,8$ and $2g/mol$. Momentum axis is in SI units.}
    \label{mass}
\end{figure}
Careful observation of the plots would reveal that there is a slight dependence on the molar mass. Slightly more active brownian motion should be expected in fluids with higher molar mass. But practically the motion hardly is effected by the composition of the fluid. This is another property of the Brownian motion as enlisted in introduction.

%% file: sections/conclusions.tex
\section{Conclusions}

The probability density function so obtained is consistent with experimental observations. At higher temperatures and small enough particle size, the probability density function for the unbalanced force tends to become completely uniform and thus making the impulse random. But for low enough temperatures, as one can see from Figure {\ref{temp}} or even for large enough particle size, as one can see in Figure {\ref{size}}, one should observe a more predictable motion of the particle governed by a more predictable(non-uniform) density function of the unbalanced force. The time average of the unbalanced force, $<xF(t)>$ in Langevin's equation can here be replaced by the ensemble average (average over all possible subsets that could have collided with the particle at a particular time). Here, $<A>$ represents the average of $A$. Ensemble average is nothing but equal to the average of $xMv$ calculated using the PDF $\Phi(Mv)$. Thus the Langevin's equation can be now solved as \cite{Coffey},
$$- m<\left(\frac{dx}{dt}\right)^2>  + \frac{m}{2}\frac{d}{dt}\left(\frac{<x^2>}{dt}\right) = -\frac{\zeta}{2}\frac{d<x^2>}{dt} + <Fx>$$
Rather than taking the last term on right hand side of the above equation to be vanishing, for general case, one should write it as,
\begin{equation}
  - m<\left(\frac{dx}{dt}\right)^2> + \frac{m}{2}\frac{d}{dt}\left(\frac{<x^2>}{dt}\right)  = -\frac{\zeta}{2}\frac{d<x^2>}{dt} + \eta(x)
\end{equation}
where $\eta(x)$ is the calculated average of $xMv$ using $\Phi(Mv)$. Solving this differential equation will give the general solution for displacement of the brownian particle, $x(t)$. 

\begin{itemize}
    \item I discover that at low temperatures(lower than $\sim 3K$), the unbalanced impulse causing the Brownian motion no longer behaves as a white noise, but obeys a probability distribution function.
    \item At higher temperatures, smaller particle sizes, the probability distribution unbalanced impulse can be  range through large values will almost equal probabilities.
\end{itemize}

%% file: sections/Acknowledgement.tex
\section{Acknowledgement}

The author is thankful to Priyanka Samanta and Bidhi Vijaywargia for their crucial insights which helped the author overcome the plateau periods. The author is also grateful to Venkat Sai Reddy and Snehil Pandey for their valuable discussions. The author is grateful to Dr.Jayeeta Chowdhury for reviewing an earlier version of this work and showing encouragement to complete it.